\documentclass[10pt, conference]{IEEEtran}
\usepackage[dvips]{graphicx}
\usepackage{graphics, subfigure, color, epsfig, psfrag, amsmath, amsfonts, amssymb, eucal, balance}
\usepackage{bm}
\usepackage{algorithm}
\usepackage{algorithmic}
\usepackage{geometry}

\newtheorem{cor}{Corollary}

\newtheorem{prop}{Proposition}

\begin{document}
\title{A Mixed-ADC Receiver Architecture for Massive MIMO Systems}
\author{\authorblockN{Ning Liang and Wenyi Zhang}\\
\authorblockA{
Key Laboratory of Wireless-Optical Communications, Chinese Academy of Sciences\\
Department of EEIS, University of Science and Technology of China\\
Emails: {\tt liangn@mail.ustc.edu.cn, wenyizha@ustc.edu.cn}}}\vspace{-10pt}
\maketitle
\thispagestyle{empty}
\begin{abstract}
Motivated by the demand for energy-efficient communication solutions in the next generation cellular network, a mixed-ADC receiver architecture for massive multiple input multiple output (MIMO) systems is proposed, which differs from previous works in that herein one-bit analog-to-digital converters (ADCs) partially replace the conventionally assumed high-resolution ADCs. The information-theoretic tool of generalized mutual information (GMI) is exploited to analyze the achievable data rates of the proposed system architecture and an array of analytical results of engineering interest are obtained. For deterministic single input multiple output (SIMO) channels, a closed-form expression of the GMI is derived, based on which the linear combiner is optimized. Then, the asymptotic behaviors of the GMI in both low and high SNR regimes are explored, and the analytical results suggest a plausible ADC assignment scheme. Finally, the analytical framework is applied to the multi-user access scenario, and the corresponding numerical results demonstrate that the mixed system architecture with a relatively small number of high-resolution ADCs is able to achieve a large fraction of the channel capacity without output quantization.
\end{abstract}
\setcounter{page}{1}

\section{Introduction}
The prosperity of mobile Internet calls for new technologies to meet the exponential increase in demand for mobile data traffic. In recent years, a heightened attention has been focused on massive multiple input multiple output (MIMO) systems, which achieves significant gains in both energy efficiency and spectral efficiency, and thus are envisioned as a promising key enabler for the next generation cellular network \cite{marzetta2010noncooperative} \cite{ngo2013energy}.

Thus far, most of the works on massive MIMO assume perfect hardware implementation. However, this assumption is not well justified, since the hardware cost and circuit power consumption scale linearly with the number of BS antennas and thus soon become economically unbearable unless low-cost, energy-efficient hardware is deployed which however easily suffers from impairments. Among various sources of hardware impairment, low-resolution analog-to-digital converters (ADCs) have attracted ubiquitous attention due to their favorable property of low cost, low power consumption and feasibility of implementation \cite{le2005analog}. For Nyquist-sampled real Gaussian channel, the authors of \cite{singh2009limits} established some general results regarding low-resolution output quantization. The authors of \cite{mezghani2007modified} designed a modified minimum mean square error (MMSE) receiver for MIMO systems with output quantization. In \cite{yin2010monobit}, the authors investigated a practical monobit digital receiver paradigm for impulse radio ultra-wideband (UWB) systems. Recently, the authors of \cite{risi2014massive} examined the impact of one-bit quantization on achievable rates of massive MIMO systems with both perfect and estimated channel state information (CSI). The authors of \cite{mo2014high} addressed the high signal-to-noise (SNR) capacities of both single input multiple output (SIMO) and MIMO channels with one-bit quantization.

Despite of its great superiority in deployment cost and energy efficiency, one-bit quantization generally has to tolerate large rate loss, especially in the high SNR regime \cite{mo2014high}, thus highlighting the indispensability of high-resolution ADC for digital receiver. Motivated by such consideration, in this paper we propose a mixed receiver architecture for massive MIMO systems in which one-bit ADCs partially, but not completely, replace conventionally assumed high-resolution ADCs. This architecture has the potential of allowing us to remarkably reduce the hardware cost and power consumption while still maintain a large fraction of the performance gains promised by massive MIMO.

Recognizing the challenge in working with the channel capacity directly, we take an alternative route and seek to characterize the achievable data rates specified to certain encoding/decoding scheme. To this end, we exploit the information-theoretic tool of generalized mutual information (GMI) \cite{lapidoth2002fading} to address the achievable data rates of our proposed system architecture. As a performance measure for mismatched decoding, GMI has proved convenient and useful in several important scenarios such as fading channels with imperfect CSI at the receiver \cite{lapidoth2002fading} and channels with transceiver distortion \cite{zhang2012general}, \cite{vehkapera2015asymptotic}.

Exploiting a general analytical framework developed in \cite{zhang2012general}, we obtain a series of analytical results. First, we consider a deterministic SIMO channel where the BS is equipped with $N$ antennas but only has access to $K$ pairs of high-resolution ADCs and $(N-K)$ pairs of one-bit ADCs, and derive a closed-form expression of the GMI for general ADC assignment and linear combiner. This enables us to optimize the linear combiner design and further explore the asymptotic behaviors of the GMI in both low and high SNR regimes that in turn help us suggest a plausible ADC assignment scheme. The corresponding numerical results indicate that even with a small number of high-resolution ADCs, our system architecture already achieves a substantial fraction of the channel capacity without output quantization, thus verifying the effectiveness of the mixed architecture. Further issues such as ergodic fading channels with or without channel estimation error are included in an extended work \cite{liang2015mixed}.

Then, we apply our analysis to the multi-user access scenario. Numerical results reveal that the mixed system architecture with a small number of high-resolution ADCs achieves a large fraction of the channel capacity without output quantization, provided that the multi-user system is properly loaded. In summary, the proposed mixed architecture strikes a reasonable and attractive balance between cost and spectral efficiency, for both single-user and multi-user scenarios. Thus we envision it as a promising receiver paradigm for energy-efficient massive MIMO systems.

The remaining part of this paper is organized as follows. Section \ref{sect:system architecture} outlines the system model. Adopting GMI as the performance metric, Section \ref{sect:GMI and combiner} establishes the theoretical framework for deterministic SIMO channels, based on which the optimal linear combiner design and the asymptotic behaviors of the GMI in both low and high SNR regimes are explored. Section \ref{sect:multi-user} extends the theoretical framework to the multi-user access scenario. Numerical results are given in Section \ref{sect:numerical} to corroborate the analysis. Finally, Section \ref{sect:conclusion} concludes the paper.

\emph{Notation:} Throughout this paper, we use $\|\mathbf{x}\|$ to represent the 2-norm of vector $\mathbf{x}$, and let $\mathbf{X}^{*}$, $\mathbf{X}^{T}$ and $\mathbf{X}^{H}$ denote the conjugate, transpose and conjugate transpose of $\mathbf{X}$, respectively. Complex Gaussian distribution with mean $\mu$ and variance $\sigma^2$ is denoted by $\mathcal{CN}(\mu,\sigma^2)$, while $\mathcal{CN}(\bm{\mu},\mathbf{C})$ stands for the distribution of a circularly symmetric complex Gaussian random vector with mean $\bm{{\mu}}$ and covariance matrix $\mathbf{C}$. We use $\log(x)$ to denote the natural logarithm of positive real number $x$.

\section{System model}
\label{sect:system architecture}
We start by focusing on a single-user system where a single-antenna user communicates with an $N$-antenna BS. In this paper, we consider a narrow-band channel model, for which the frequency-flat fading channel $\mathbf{h}$ is chosen according to $\mathcal{CN}(\mathbf{0},\mathbf{I})$ and is fixed throughout the transmission of the codeword. Moreover, the realization of the channel is assumed to be perfectly known by the BS and thus is deemed as deterministic in the subsequent analysis. The received signal at the BS can be expressed as
\begin{equation}
\mathbf{y}^l=\mathbf{h}x^l+\mathbf{z}^l,\ \ \ \mathrm{for}\  l=1,2,...,L,
\label{equ:equ_1}
\end{equation}
where $x^l$ is the complex signal transmitted at the $l$-th symbol time, $\mathbf{z}^l\sim\mathcal{CN}(\mathbf{0},\sigma^2\mathbf{I})$ models the independent and identically distributed (i.i.d.) complex Gaussian noise vector, and $L$ is the codeword length.

\begin{figure}
\centering
\includegraphics[width=0.48\textwidth]{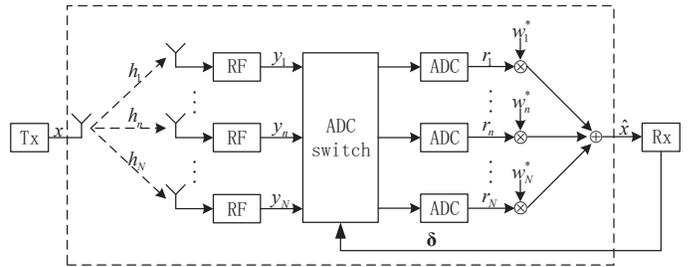}
\centering
\caption{Illustration of the system architecture.} \label{fig:SystemModel}
\end{figure}

We consider a mixed architecture, as illustrated in Figure \ref{fig:SystemModel}, in which only $2K$ high-resolution ADCs are available and all the other $2(N-K)$ ADCs are with only one-bit resolution. We further let the I/Q outputs at each antenna be quantized by two ADCs of the same kind. Thus the quantized output can be expressed as
\begin{equation}
r_{n}^{l}=\delta_{n}\cdot(h_{n} x^{l}+z_{n}^{l})+\bar{\delta}_n\cdot\mathrm{sgn}(h_{n} x^{l}+z_{n}^{l}),
\label{equ:equ_3}
\end{equation}
for $l=1,...,L,\ n=1,...,N$. Here $\bar{\delta}_n\triangleq 1-\delta_n$, and $\delta_n\in\{0,1\}$ is the binding indicator: $\delta_n=1$ means that the ADCs corresponding to the $n$-th antenna are high-resolution, whereas $\delta_n=0$ indicates that they are with one-bit resolution. Note that here we assume sufficiently high resolution, so that the residual quantization noise is negligible, for $\delta_n=1$. For notation simplicity, we further define a binding vector $\bm{\delta}\triangleq[\delta_1,...,\delta_N]^{T}$,
which should be optimized according to the channel state $\mathbf{h}$ so that the limited number of high-resolution ADCs will be well utilized to enhance the system performance.

For transmission of rate $R$, the user selects a message $m$ from $\mathcal{M}=\{1,2,...,\lfloor 2^{LR} \rfloor\}$ uniformly randomly, and maps the selected message to a transmitted codeword, i.e., a length-$L$ complex sequence, $\{x^l(m)\}_{l=1}^{L}$. In this paper, we restrict the codebook to be drawn from a Gaussian ensemble; that is, each codeword is a sequence of $L$ i.i.d. $\mathcal{CN}(0,\mathcal{E}_\mathrm{s})$ random variables, and all the codewords are mutually independent. Such a choice of codebook satisfies the average power constraint $\frac{1}{L}\sum_{l=1}^{L}\mathbb{E}[|x^{l}(m)|^2]\leq\mathcal{E}_\mathrm{s}$. We thus define the SNR as $\mathrm{SNR}_{\mathrm{d}}=\mathcal{E}_\mathrm{s}/\sigma^2$, and let $\sigma^2=1$ thereafter for convenience.

For an $N$-antenna SIMO channel, we let $C(N,K_1,K_2)$ denote its capacity when equipped with $K_1$ pairs of high-resolution ADCs and $K_2$ pairs of one-bit ADCs, where $0\leq K_1, K_2, K_1+K_2 \leq N$. As discussed in the introduction, the evaluation of $C(N,K,N-K)$ is a formidable task. Therefore in the following, we adopt the nearest-neighbor decoding rule at the decoder, and leverage the general framework developed in \cite{zhang2012general} to investigate the GMI of our proposed system architecture. The GMI acts as an achievable rate and thus also a lower bound of $C(N,K,N-K)$. To this end, we introduce a linear combiner to process the channel output vector, as illustrated in Figure \ref{fig:SystemModel}. Thus the processed channel output is
\begin{equation}
\hat{x}^{l}=\mathbf{w}^{H}\mathbf{r}^{l},
\label{equ:equ_6}
\end{equation}
for $l=1,...,L$, where $\mathbf{w}$ is designed according to the channel state $\mathbf{h}$ and the binding vector $\bm{\delta}$.

With nearest-neighbor decoding, upon observing $\{\hat{x}^{l}\}_{l=1}^L$, the decoder computes, for all possible messages, the distance metrics
\begin{equation}
D(m)=\frac{1}{L}\sum_{l=1}^{L}|\hat{x}^{l}-a x^{l}(m)|^2,\ \ \ m\in\mathcal{M},
\label{equ:equ_7}
\end{equation}
and decides the received message as the one that minimizes \eqref{equ:equ_7}. The scaling parameter $a$ is selected appropriately for optimizing the decoding performance.

\section{GMI and Optimization of Combiner}
\label{sect:GMI and combiner}
\subsection{GMI of the Proposed System Architecture}
\label{subsect:GMI}
From now on, we suppress the time index $l$ for notational simplicity. To facilitate the exposition, we summarize \eqref{equ:equ_3} and \eqref{equ:equ_6} as
\begin{equation}
\hat{x}=\mathbf{w}^{H}\mathbf{r}\triangleq f(x,\mathbf{h},\mathbf{z}),
\label{equ:equ_8}
\end{equation}
where $f(\cdot)$ is a memoryless nonlinear distortion function that incorporates the effects of output quantization as well as linear combining. Although $\bm{\delta}$ and $\mathbf{w}$ are made invisible in the function $f(\cdot)$ since they are both determined by $\mathbf{h}$, we need to keep in mind that $f(\cdot)$ implicitly includes $\bm{\delta}$ and $\mathbf{w}$.

We apply the general framework developed in \cite{zhang2012general} to derive the GMI of our proposed system architecture. Particularly, employing the similar procedure as \cite[Appendix C]{zhang2012general}, we have the conditional GMI given as follows.

\begin{prop}
\label{prop:prop_1}
With Gaussian codebook ensemble and nearest-neighbor decoding, the GMI conditioned on $\mathbf{w}$ and $\bm{\delta}$ is given as
\begin{equation}
I_{\mathrm{GMI}}(\mathbf{w},\bm{\delta})=\mathrm{log}\left(1+\frac{\kappa(\mathbf{w},\bm{\delta})}{1-\kappa(\mathbf{w},\bm{\delta})}\right),
\label{equ:equ_9}
\end{equation}
where the parameter $\kappa(\mathbf{w},\bm{\delta})$ is
\begin{equation}
\kappa(\mathbf{w},\bm{\delta})=\frac{|\mathbb{E}[f^{*}(x,\mathbf{h},\mathbf{z})\cdot x]|^2}
{\mathcal{E}_\mathrm{s} \mathbb{E}[|f(x,\mathbf{h},\mathbf{z})|^2]}.
\label{equ:equ_10}
\end{equation}
The corresponding optimal choice of the scaling parameter $a$ is
\begin{equation}
a_{\mathrm{opt}}(\mathbf{w},\bm{\delta})=\frac{\mathbb{E}[f(x,\mathbf{h},\mathbf{z})\cdot x^{*}]}{\mathcal{E}_\mathrm{s}}.
\label{equ:equ_11}
\end{equation}
\end{prop}

Note that $\kappa(\mathbf{w},\bm{\delta})$ is the squared correlation coefficient of channel input $x$ and the processed output $f(x,\mathbf{h},\mathbf{z})$, and thus is upper bounded by one, from Cauchy-Schwartz's inequality. Moreover, $I_{\mathrm{GMI}}(\mathbf{w},\bm{\delta})$ is a strictly increasing function of $\kappa(\mathbf{w},\bm{\delta})$ for $\kappa(\mathbf{w},\bm{\delta})\in(0,1)$. Therefore in the following, we will seek to maximize $\kappa(\mathbf{w},\bm{\delta})$ by choosing well designed linear combiner $\mathbf{w}$ and binding vector $\bm{\delta}$. To start with, we derive a closed-form expression of $\kappa(\mathbf{w},\bm{\delta})$, given as follows.

\begin{prop}
\label{prop:prop_2}
Given $\mathbf{w}$ and $\bm{\delta}$, for \eqref{equ:equ_10} in Proposition \ref{prop:prop_1}, we have
\begin{equation}
\kappa(\mathbf{w},\bm{\delta})=\frac{\mathbf{w}^{H}\mathbf{R}_{\mathbf{r}x}\mathbf{R}_{\mathbf{r}x}^{H}\mathbf{w}}
{\mathcal{E}_\mathrm{s}\mathbf{w}^{H}\mathbf{R}_{\mathbf{rr}}\mathbf{w}},
\label{equ:equ_17}
\end{equation}
where $\mathbf{R}_{\mathbf{r}x}\triangleq\mathbb{E}[\mathbf{r}x^*]$ is the correlation vector between $\mathbf{r}$ and $x$, with its $n$-th element being
\begin{equation}
(\mathbf{R}_{\mathbf{r}x})_n=h_n \mathcal{E}_\mathrm{s} \left[\delta_n+\bar{\delta}_n\cdot\sqrt{\frac{4}{\pi(|h_n|^2\mathcal{E}_\mathrm{s}+1)}}\right],
\label{equ:equ_18}
\end{equation}
and $\mathbf{R}_{\mathbf{rr}}\triangleq\mathbb{E}[\mathbf{r}\mathbf{r}^H]$ is the covariance matrix of $\mathbf{r}$, with its $(n,m)$-th entry being $(\mathbf{R}_{\mathbf{rr}})_{n,m}=$
\begin{equation}
\begin{cases}
1+\delta_n\cdot |h_n|^2\mathcal{E}_\mathrm{s}+\bar{\delta}_n,&\mathrm{if}\ n=m, \\
h_nh_m^*\mathcal{E}_\mathrm{s}\Bigg[\delta_n\delta_m+
\delta_n\bar{\delta}_m\cdot\sqrt{\frac{4}{\pi(|h_m|^2\mathcal{E}_\mathrm{s}+1)}}+\\
\ \ \ \ \ \ \ \ \ \ \ \bar{\delta}_n\delta_m\cdot\sqrt{\frac{4}{\pi(|h_n|^2\mathcal{E}_\mathrm{s}+1)}}\Bigg]+\\
\bar{\delta}_n\bar{\delta}_m\!\cdot\!\frac{4}{\pi}\Bigg[
\mathrm{arcsin}\Big(\frac{(h_nh_m^*)^{\mathrm{R}}\mathcal{E}_\mathrm{s}}
{\sqrt{|h_n|^2\mathcal{E}_\mathrm{s}+1}\sqrt{|h_m|^2\mathcal{E}_\mathrm{s}+1}}\Big)+\\
\ \ \ \ \ \ \ \ \ \ \ i\!\cdot\!\mathrm{arcsin}\Big(\frac{(h_nh_m^*)^{\mathrm{I}}\mathcal{E}_\mathrm{s}}
{\sqrt{|h_n|^2\mathcal{E}_\mathrm{s}+1}\sqrt{|h_m|^2\mathcal{E}_\mathrm{s}+1}}\Big)
\Bigg],&\mathrm{if}\ n\neq m.
\end{cases}
\label{equ:equ_19}
\end{equation}
The corresponding optimal choice of the scaling parameter $a$ in \eqref{equ:equ_11} is
\begin{equation}
a_{\mathrm{opt}}(\mathbf{w},\bm{\delta})=\frac{1}{\mathcal{E}_\mathrm{s}}\mathbf{w}^{\mathrm{H}}\mathbf{R}_{\mathbf{r}x}.
\label{equ:equ_20}
\end{equation}
\end{prop}

\emph{Proof:} See \cite[Appendix A]{liang2015mixed}.
\subsection{Optimization of Linear Combiner}
\label{subsect:combiner}
In this subsection, we turn to optimize $\mathbf{w}$ such that the GMI is maximized for given $\mathbf{h}$ and $\bm{\delta}$. The subsequent proposition summarizes our result.

\begin{prop}
\label{prop:prop_3}
For given $\mathbf{h}$ and $\bm{\delta}$, the optimal linear combiner $\mathbf{w}$ takes the following form
\begin{equation}
\mathbf{w}_{\mathrm{opt}}=\mathbf{R}_{\mathbf{rr}}^{-1}\mathbf{R}_{\mathbf{r}x},
\label{equ:equ_21}
\end{equation}
which is in fact a minimum mean square error (MMSE) combiner that minimizes the mean squared estimation error of $x$ upon observing $\mathbf{r}$. The corresponding $\kappa(\mathbf{w},\bm{\delta})$ is
\begin{equation}
\kappa(\mathbf{w}_{\mathrm{opt}},\bm{\delta})=a_{\mathrm{opt}}(\mathbf{w}_{\mathrm{opt}},\bm{\delta})=
\frac{1}{\mathcal{E}_\mathrm{s}}\mathbf{R}_{\mathbf{r}x}^{H}\mathbf{R}_{\mathbf{rr}}^{-1}\mathbf{R}_{\mathbf{r}x}.
\label{equ:equ_22}
\end{equation}
\end{prop}

\emph{Proof:} See \cite[Section III-B]{liang2015mixed} for the proof.
\subsection{Asymptotic Behaviors of $I_{\mathrm{GMI}}(\mathbf{w}_{\mathrm{opt}},\bm{\delta})$}
\label{subsect:asymptotic behaviors}
In the previous subsection, the optimal linear combiner for our proposed framework is derived. Thus we are ready to examine its asymptotic performance in both low and high SNR regimes. Letting SNR tend to zero, we have the following corollary.

\begin{cor}
\label{cor:cor_2}
As $\mathcal{E}_\mathrm{s}\rightarrow 0$, for given $\bm{\delta}$ we have
\begin{equation}
I_{\mathrm{GMI}}(\mathbf{w}_{\mathrm{opt}},\bm{\delta})
=\sum_{n=1}^{N}\left(\delta_n+\bar{\delta}_n\cdot\frac{2}{\pi}\right)|h_n|^2\mathcal{E}_\mathrm{s}+o(\mathcal{E}_\mathrm{s}).
\label{equ:equ_28}
\end{equation}
\end{cor}

See \cite[Appendix B]{liang2015mixed} for its proof. Comparing with $C(N,N,0)$ in the low SNR regime, i.e., $C(N,N,0)$ $=\sum_{n=1}^{N} |h_n|^2\mathcal{E}_\mathrm{s}+o(\mathcal{E}_\mathrm{s})$, we conclude that part of the achievable rate is degraded by a factor of $\frac{2}{\pi}$ due to one-bit quantization.

For the high SNR case, the subsequent corollary collects our results.

\begin{cor}
\label{cor:cor_3}
As $\mathcal{E}_\mathrm{s}\rightarrow\infty$, for given $\bm{\delta}$ we have the effective SNR in \eqref{equ:equ_9} as
\begin{equation}
\frac{\kappa(\mathbf{w}_{\mathrm{opt}},\bm{\delta})}{1-\kappa(\mathbf{w}_{\mathrm{opt}},\bm{\delta})}=
\|\mathbf{p}\|^2\mathcal{E}_\mathrm{s}
+\frac{[4+O(1/\mathcal{E}_\mathrm{s})]\mathbf{q}^{H}\mathbf{B}^{-1}\mathbf{q}}
{\pi-[4+O(1/\mathcal{E}_\mathrm{s})]\mathbf{q}^{H}\mathbf{B}^{-1}\mathbf{q}},
\label{equ:equ_29}
\end{equation}
with $\mathbf{p}$, $\mathbf{q}$, and $\mathbf{B}$ given by
\begin{eqnarray}
\mathbf{p}&\!\!\!\!\!\!\!\!\triangleq\!\!\!\!\!\!\!\!&[\underline{h}_1,...,\underline{h}_K]^{T},\\
\mathbf{q}&\!\!\!\!\!\!\!\!\triangleq\!\!\!\!\!\!\!\!&\left[\underline{h}_{K+1}/|\underline{h}_{K+1}|,...,\underline{h}_{N}/|\underline{h}_{N}|\right]^{T},\\
(\mathbf{B})_{n,m}&\!\!\!\!\!\!\!\!\triangleq\!\!\!\!\!\!\!\!&\frac{4}{\pi}\Bigg[\arcsin\left(\frac{(\underline{h}_{n+K}\underline{h}_{m+K}^*)^{\mathrm{R}}}
{|\underline{h}_{n+K}\underline{h}_{m+K}^*|}\right)+\nonumber\\
&&i\cdot\arcsin\left(\frac{(\underline{h}_{n+K}\underline{h}_{m+K}^*)^{\mathrm{I}}}
{|\underline{h}_{n+K}\underline{h}_{m+K}^*|}\right)\Bigg]\!+\!O(1/\mathcal{E}_\mathrm{s}),
\end{eqnarray}
where $\underline{\mathbf{h}}$ is a rearrangement of $\mathbf{h}$, by stacking the channel coefficients of the antennas equipped with high-resolution ADCs in the first $K$ positions of $\underline{\mathbf{h}}$.
\end{cor}

See \cite[Appendix C]{liang2015mixed} for the proof. An observation from \eqref{equ:equ_29} indicates that the contributions of high-resolution ADCs and one-bit ADCs in the high SNR regime are separate, as the first term corresponding to high-resolution ADCs increases linearly with $\mathcal{E}_\mathrm{s}$, whereas the second term coming from one-bit ADCs tends to be a positive constant independent of $\mathcal{E}_\mathrm{s}$. Comparing with Corollary \ref{cor:cor_2}, we infer that one-bit ADCs are getting less beneficial as the SNR grows large, as will be validated by numerical study in Section \ref{sect:numerical}. Moreover, both \eqref{equ:equ_28} and \eqref{equ:equ_29} suggest that high-resolution ADCs should be assigned to the antennas with the strongest $K$ link magnitude gains, at least in the low and high SNR regimes.

\section{Extension to multi-user scenario}
\label{sect:multi-user}
In this section, the BS serves $M$ single-antenna users simultaneously. For simplicity, we focus on the deterministic channel case, and the channel matrix between the users and the BS is denoted by $\mathbf{H}\triangleq[\mathbf{h}_1,...,\mathbf{h}_N]\in\mathbb{C}^{M\times N}$, whose elements are i.i.d. $\mathcal{CN}(0,1)$, i.e., $\mathbf{h}_n\triangleq[h_{1n},...,h_{Mn}]^T$ collecting the channel coefficients related to the $n$-th BS antenna. There are still only $K$ pairs of high-resolution ADCs available at the BS. Thus we rewrite the quantized output at the $n$-th antenna, with user $j$ considered, as
\begin{eqnarray}
r_n^{\mathrm{mu}}\!=\!\delta_n\!\cdot\!\left(\sum_{\iota=1}^{M}h_{\iota n}x_\iota\!+\!z_n\right)\!+\!
\bar{\delta}_n\!\cdot\!\mathrm{sgn}\left(\sum_{\iota=1}^{M}h_{\iota n}x_\iota\!+\!z_n\right),
\label{equ:equ_43}
\end{eqnarray}
where $x_{\iota}\sim\mathcal{CN}(0,\mathcal{E}_\mathrm{s})$ denotes the i.i.d. coded signal dedicated to the $\iota$-th user, and $\sum_{\iota\neq j}^{M}h_{\iota n}x_\iota+z_n$ summarizes the co-channel interference and noise for the considered user $j$. For a fair comparison, the SNR in this situation is defined as $\mathrm{SNR}_{\mathrm{d}}=M\mathcal{E}_\mathrm{s}$, reflecting the total transmit power from all users.

Following a similar derivation procedure as that constructed in Section \ref{sect:GMI and combiner}, we get the GMI of the considered user.

\begin{prop}
\label{prop:prop_5}
For given $\mathbf{H}$ and $\bm{\delta}$, when treating other users' signals as noise, the GMI of user $j$ is
\begin{equation}
I_{\mathrm{GMI}}^{\mathrm{mu}}=\log\left(1+\frac{\kappa^{\mathrm{mu}}}{1-\kappa^{\mathrm{mu}}}\right),
\label{equ:equ_44}
\end{equation}
where the parameter $\kappa^{\mathrm{mu}}$ is
\begin{equation}
\kappa^{\mathrm{mu}}=\frac{1}{\mathcal{E}_\mathrm{s}}
(\mathbf{R}_{\mathbf{r}x}^{\mathrm{mu}})^{H}(\mathbf{R}_{\mathbf{rr}}^{\mathrm{mu}})^{-1}\mathbf{R}_{\mathbf{r}x}^{\mathrm{mu}}.
\label{equ:equ_45}
\end{equation}
$\mathbf{R}_{\mathbf{r}x}^{\mathrm{mu}}$ is the correlation vector between $\mathbf{r}^{\mathrm{mu}}$ and $x_j$, with its $n$-th entry given as
\begin{equation}
(\mathbf{R}_{\mathbf{r}x}^{\mathrm{mu}})_n=h_{jn} \mathcal{E}_\mathrm{s} \left[\delta_n+\bar{\delta}_n\cdot\sqrt{\frac{4}{\pi(\|\mathbf{h}_n\|^2\mathcal{E}_\mathrm{s}+1)}}\right],
\label{equ:equ_46}
\end{equation}
and $\mathbf{R}_{\mathbf{rr}}^{\mathrm{mu}}$ is the covariance matrix of $\mathbf{r}^{\mathrm{mu}}$, with its $(n,m)$-th entry being $(\mathbf{R}_{\mathbf{rr}}^{\mathrm{mu}})_{n,m}=$
\begin{equation}
\begin{cases}
1+\delta_n\cdot\|\mathbf{h}_n\|^2\mathcal{E}_\mathrm{s}+\bar{\delta}_n,&\mathrm{if}\ n=m, \\
\mathbf{h}_n^T\mathbf{h}_m^* \mathcal{E}_\mathrm{s}
\Bigg[\delta_n\delta_m+\delta_n\bar{\delta}_m\cdot\sqrt{\frac{4}{\pi(\|\mathbf{h}_m\|^2\mathcal{E}_\mathrm{s}+1)}}+\\
\ \ \ \ \ \ \ \ \ \ \ \ \bar{\delta}_n\delta_m\cdot\sqrt{\frac{4}{\pi(\|\mathbf{h}_n\|^2\mathcal{E}_\mathrm{s}+1)}}\Bigg]+\\
\bar{\delta}_n\bar{\delta}_m\!\cdot\!\frac{4}{\pi}\Bigg[
\mathrm{arcsin}\Big(\frac{(\mathbf{h}_n^T\mathbf{h}_m^*)^{\mathrm{R}}\mathcal{E}_\mathrm{s}}
{\sqrt{\|\mathbf{h}_n\|^2\mathcal{E}_\mathrm{s}+1}\sqrt{\|\mathbf{h}_m\|^2\mathcal{E}_\mathrm{s}+1}}\Big)+\\
\ \ \ \ \ \ \ \ \ \ \ i\!\cdot\!\mathrm{arcsin}\Big(\frac{(\mathbf{h}_n^T\mathbf{h}_m^*)^{\mathrm{I}}\mathcal{E}_\mathrm{s}}
{\sqrt{\|\mathbf{h}_n\|^2\mathcal{E}_\mathrm{s}+1}\sqrt{\|\mathbf{h}_m\|^2\mathcal{E}_\mathrm{s}+1}}\Big)
\Bigg],&\mathrm{if}\ n\neq m.
\end{cases}
\label{equ:equ_47}
\end{equation}
\end{prop}

In the multi-user scenario, there is no clear clue about how to assign the high-resolution ADCs. For this reason, we consider two heuristic ADC assignment schemes in the numerical study. That is
\begin{itemize}
\item Scheme \#1: high-resolution ADCs are assigned to the antennas with the maximum $\sum_{j=1}^{M}|h_{jn}|^2$.
\item Scheme \#2: high-resolution ADCs are assigned randomly.
\end{itemize}
Scheme \#1 follows from the asymptotic property of $I_{\mathrm{GMI}}^{\mathrm{mu}}$ in the low SNR regime, and the proof is omitted due to space limitation.
\section{Numerical Results}
\label{sect:numerical}
In this section, we validate our previous analysis with numerical results. The GMI and the channel capacity are both averaged over a large number of channel realizations, for both single-user and multi-user scenarios.
\subsection{GMI for Single-User Scenario}
\begin{figure}
\centering
\includegraphics[width=0.4\textwidth]{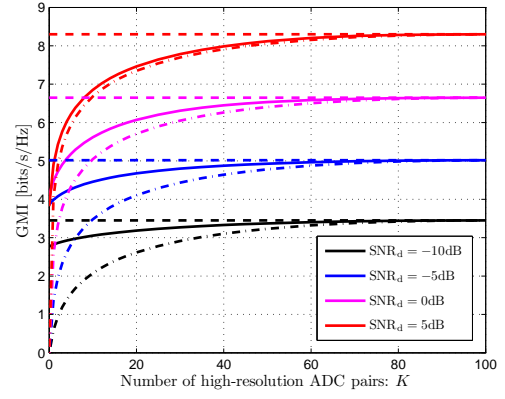}
\caption{GMI of the proposed system architecture for different number of high-resolution ADCs, $N=100$.}
\label{fig:fig_2}
\end{figure}

\begin{figure}
\centering
\includegraphics[width=0.4\textwidth]{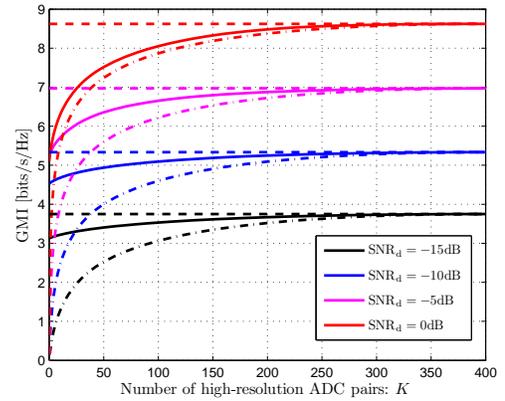}
\caption{GMI of the proposed system architecture for different number of high-resolution ADCs, $N=400$.}
\label{fig:fig_3}
\end{figure}

We assign the high-resolution ADCs to the antennas with the strongest $K$ link magnitude gains, as suggested by Corollary \ref{cor:cor_2} and Corollary \ref{cor:cor_3}. In Figure \ref{fig:fig_2} and Figure \ref{fig:fig_3}, the solid curves represent $I_{\mathrm{GMI}}(\mathbf{w}_{\mathrm{opt}},\bm{\delta})$, the dashed lines correspond to the channel capacity without output quantization, i.e., $C(N,N,0)=\mathrm{log}(1+\|\mathbf{h}\|^2\mathrm{SNR}_{\mathrm{d}})$, and the dash-dot curves account for the channel capacity under antenna selection, i.e., $C(N,K,0)=\mathrm{log}(1+\sum_{n=1}^N \delta_n\cdot|h_n|^2\mathrm{SNR}_{\mathrm{d}})$. Several observations are in order. First, both figures clearly show that the gain of deploying more high-resolution ADCs decreases rapidly, and thus a small number of high-resolution ADCs actually already achieves a large fraction of $C(N,N,0)$. For example, our proposed system architecture with 10 pairs of high-resolution ADCs attains 84\% of $C(100,100,0)$ when $\mathrm{SNR}_{\mathrm{d}}=0\mathrm{dB}$, and it achieves 91\% of $C(100,100,0)$ when $K$ increases to 20. Besides, both figures indicate that, compared to the antenna selection paradigm, one-bit ADCs in our architecture are less beneficial when the SNR grows large, but still significantly improve the system performance in the low to intermediate SNR regime.
\subsection{GMI for Multi-user Scenario}
\begin{figure}
\centering
\includegraphics[width=0.4\textwidth]{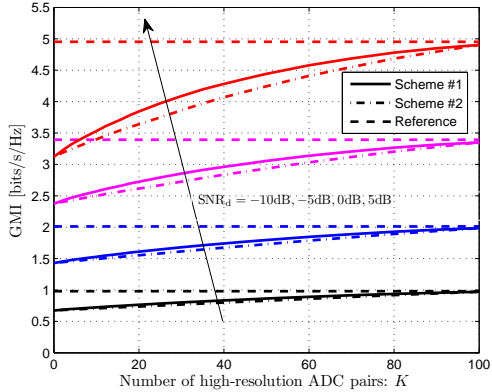}
\caption{GMI of the considered user $j$ in multi-user scenario, $N=100$, $M=10$.}
\label{fig:fig_12}
\end{figure}

Now, we examine the feasibility of our proposed system architecture in the multi-user scenario. Figure \ref{fig:fig_12} collects the result, where the solid curves correspond to Scheme \#1 suggested by asymptotic analysis in the low SNR regime, the dash-dot curves are obtained by random assignment per Scheme \#2, and the dashed lines refer to the per-user capacity with $N$ pairs of high-resolution ADCs, i.e., $\frac{1}{M}\log\det(\mathbf{I}+\mathcal{E}_\mathrm{s}\mathbf{H}\mathbf{H}^H)$.

We notice that though Scheme \#1 is only analytically validated for the low SNR case, it does achieve better performance than Scheme \#2. For the special case of $K=N$, it is well known that the linear MMSE receiver is suboptimal for MIMO channel \cite{tse2005fundamentals}, and thus we observe a distinguishable gap between the per-user capacity and the GMI. Most importantly, here a small number of high-resolution ADCs also attain a large fraction of the channel capacity with $N$ pairs of high-resolution ADCs. For example, when $\mathrm{SNR}_{\mathrm{d}}=0$dB and $N=100$, Scheme \#1 with $K=10$ achieves 76\% of the per-user capacity with $N$ pairs of high-resolution ADCs, and this number rises to 80\% when $K=20$.

\begin{figure}
\centering
\includegraphics[width=0.45\textwidth]{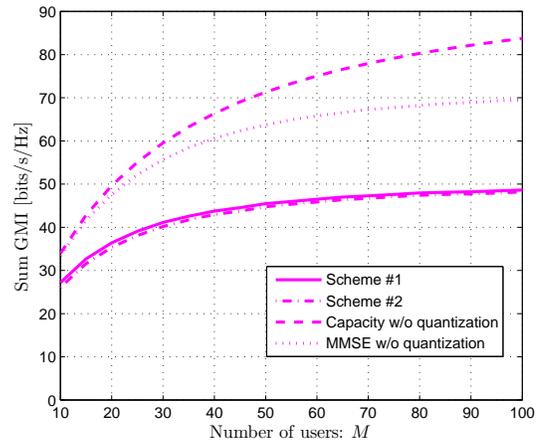}
\caption{Sum GMI for users in multi-user scenario: increasing number of users, $N=100$, $K=20$, $\mathrm{SNR}_{\mathrm{d}}$=0dB.}
\label{fig:fig_14}
\end{figure}

Figure \ref{fig:fig_14} accounts for the impact of increasing number of users on the achievable sum rates, focusing on $K=20$ and $\mathrm{SNR}_{\mathrm{d}}=0\mathrm{dB}$. For a fair comparison, we introduce another performance curve corresponding to MMSE receiver with $N$ pairs of high-resolution ADCs. We conclude that our proposed architecture achieves satisfactory performance even when the system is heavily loaded, by noticing that, even for $M=N=100$, our proposed architecture still attains 70\% of that achieved by MMSE receiver without quantization.
\section{Conclusion}
\label{sect:conclusion}
In this paper, we propose a mixed-ADC receiver architecture, and leverage the GMI to analytically evaluate its achievable data rates under two scenarios. The corresponding numerical study concludes that the proposed architecture with a small number of high-resolution ADCs suffices to achieve a significant fraction of the channel capacity without output quantization, for both single-user and multi-user scenarios. Thus our approach provides a systematic way of designing energy-efficient massive MIMO receivers.

A number of interesting and important problems remain unsolved beyond this paper, such as designing the optimal ADC assignment scheme for any SNR, especially for the multi-user scenario; extending the analysis to more comprehensive hardware impairment models beyond ADC; among others. Besides, in order to make this approach effective for wideband channels which are more prevailing in the future communication systems, it is particularly crucial to extend the analysis to frequency-selective fading channels. We note that this is feasible but beyond the scope of this paper, and will be treated in a future work.

\section*{Acknowledgement}
The research has been supported by the National Basic Research Program of China (973 Program) through grant 2012CB316004, and National Natural Science Foundation of China through grant 61379003.

\end{document}